# On the freeze quantifier in Constraint LTL: decidability and complexity ⋆

Stéphane Demri [a,1], Ranko Lazić [b,2], David Nowak [c,3]

[a] *LSV, CNRS & ENS Cachan & INRIA Futurs (projet SECSI), France*
[b] *Department of Computer Science, University of Warwick, United Kingdom*
[c] *Research Center for Information Security, National Institute of Advanced Industrial Science and Technology, Japan*

**Abstract**

Constraint LTL, a generalisation of LTL over Presburger constraints, is often used as a formal language to specify the behavior of operational models with constraints. The freeze quantifier can be part of the language, as in some real-time logics, but this variable-binding mechanism is quite general and ubiquitous in many logical languages (first-order temporal logics, hybrid logics, logics for sequence diagrams, navigation logics, logics with $\lambda$-abstraction etc.). We show that Constraint LTL over the simple domain $\langle \mathbb{N}, = \rangle$ augmented with the freeze quantifier is undecidable which is a surprising result in view of the poor language for constraints (only equality tests). Many versions of freeze-free Constraint LTL are decidable over domains with qualitative predicates and our undecidability result actually establishes $\Sigma_1^1$-completeness. On the positive side, we provide complexity results when the domain is finite (EXPSPACE-completeness) or when the formulae are flat in a sense introduced in the paper. Our undecidability results are sharp (i.e. with restrictions on the number of variables) and all our complexity characterisations ensure completeness with respect to some complexity class (mainly PSPACE and EXPSPACE).

⋆ This paper is an extended version of [1].
[1] Supported by the ACI "Sécurité et Informatique" CORTOS.
[2] Supported by an invited professorship from ENS Cachan, and by grants from the EPSRC (GR/S52759/01) and the Intel Corporation. Also affiliated to the Mathematical Institute, Serbian Academy of Sciences and Arts, Belgrade.
[3] Supported by the ACI "Sécurité et Informatique" PERSÉE and by the e-Society project of MEXT. A part of this work was done while affiliated to LSV, CNRS & ENS Cachan, and Department of Information Science, University of Tokyo.

*Preprint submitted to Information and Computation*   25 September 2018

## 1 Introduction

**Model-checking for infinite-state systems.** Temporal logics are well-studied formalisms to specify the behavior of finite-state systems and the computational complexity of the model-checking problems is nowadays well-known, see e.g. a survey in [2]. However, many systems such as communication protocols have infinitely many configurations and usually the techniques for the finite case cannot be applied directly. For numerous infinite-state systems, the model-checking problem for the linear-time temporal logic LTL can be easily shown to be undecidable (counter automata, hybrid automata and more general constraint automata [3, Chapter 6]). Actually, simpler problems such as reachability are already undecidable. However, remarkable classes of infinite-state systems admit decidable model-checking problems, such as timed automata [4] and subclasses of counter automata [5,6,7,8,9]. For instance, fragments of LTL with Presburger constraints have been shown decidable over appropriate counter automata [10,11]. In order to push further the decidability border, one way consists in considering larger classes of operational models, see e.g. [5]. Alternatively, enriching the specification language is another possibility. In the paper, we are interested in studying systematically the extensions of versions of LTL over concrete domains by the so-called freeze quantifier, and in analysing the consequences in terms of decidability and computational complexity.

**A variable-binding mechanism.** The freeze quantifier in real-time logics has been introduced by Alur and Henzinger in the logic TPTL, see e.g. [12]. The formula $x \cdot \phi(x)$ binds the variable $x$ to the time $t$ of the current state: $x \cdot \phi(x)$ is semantically equivalent to $\phi(t)$. Alternatively, in the explicit clock approach [13], there is an explicit clock variable $t$ and even though in this approach the freeze variable-binding mechanism is possible, the logical formalisms from [12] and [13] are incomparable. In this paper, we want to extend some of the decidable logics from [10,11,14] to admit the freeze quantifier: $\downarrow_{y=x} \phi(y)$ holds true at a state iff $\phi(y)$ holds true at the same state with $y$ taking the value of $x$. Here, $y$ can be in the scope of temporal operators. A crucial difference with the logics in [12,13] rests on the fact that the variable $x$ may not be monotonic. We focus on decidability and complexity issues when the language of constraints (at the atomic level of the logics) is very simple in order to isolate the effects of the freeze quantifier. We know for instance that LTL over integer periodicity constraints augmented with the freeze quantifier is EXPSPACE-complete [14].

The above-mentioned variable-binding mechanism that allows the binding of logical variables to objects is very general and it has been used in the literature for various purposes. Details will be provided along the paper (see e.g. Sections 2.2 and 5). In particular, one can see flexible variables as processes, values of the domain as resources, and the freeze quantifier and rigid variables as ways to extract and store the current resource used by a process. This view is nicely illustrated in [15] by the specification of a communication protocol. In Section 2.2, we consider the case of a process requesting memory blocks.

**Our contribution.** In the paper, we analyse decidability and complexity issues of Constraint LTL augmented with the freeze quantifier. The temporal operators we consider are restricted to the standard future-time operators 'until' and 'next' (no past-time operators). $\text{CLTL}^{\downarrow}(\mathcal{D})$ denotes such a logic over the concrete domain $\mathcal{D}$. A concrete domain is composed of a non-empty set equipped with a family of relations. The atomic formulae of $\text{CLTL}^{\downarrow}(\mathcal{D})$ are based on constraints over $\mathcal{D}$ with the ability to compare values of variables at states of bounded distance (see details in the body of the paper) as done in [16,17,11,18].

First, we show that when the underlying domain $\mathcal{D}$ is finite, $\text{CLTL}^{\downarrow}(\mathcal{D})$ satisfiability is in EXPSPACE. If moreover $\mathcal{D}$ has at least two elements with the equality predicate, then $\text{CLTL}^{\downarrow}(\mathcal{D})$ is EXPSPACE-hard. As a corollary, $\text{CLTL}^{\downarrow}(D, =)$ satisfiability is EXPSPACE-complete when $|D| \geq 2$ and $D$ is finite (Section 3.2). This witnesses an exponential blow-up since satisfiability for the freeze-free fragment $\text{CLTL}(\mathcal{D})$ when $\mathcal{D}$ is finite can be easily shown in PSPACE as plain LTL [19].

When the domain $D$ is infinite, we show that $\text{CLTL}^{\downarrow}(D, =)$ is undecidable which is the main result of the paper (Section 4). This is quite surprising since the language of constraints is poor (only equality tests) and only future-time operators are used unlike what is shown in [14, Section 7] with past-time operators. Our proof, based on a reduction from the Recurrence Problem for 2-counter machines, refines this result: $\text{CLTL}^{\downarrow}(D, =)$ is $\Sigma_1^1$-complete even if only one flexible variable and two rigid variables (used to record the values of flexible variables) are involved. Hence, in spite of the very basic Presburger constraints in $\text{CLTL}^{\downarrow}(\mathbb{N}, =)$, satisfiability is $\Sigma_1^1$-complete. Decidability of $\text{CLTL}^{\downarrow}(\mathcal{D})$ can be obtained either at the cost of syntactic restrictions or by assuming semantical constraints (as in the logic TPTL [12] where the freeze quantifier can only record the value of a monotonic variable, namely time).

In order to regain decidability, we introduce the flat fragment of $\text{CLTL}^{\downarrow}(\mathcal{D})$ which contains the freeze-free fragment $\text{CLTL}(\mathcal{D})$ and we show that there is a logarithmic-space reduction from the flat fragment of $\text{CLTL}^{\downarrow}(\mathcal{D})$ into $\text{CLTL}(\mathcal{D})$ assuming that the equality predicate belongs to $\mathcal{D}$. As a corollary, we obtain that the flat fragments of $\text{CLTL}^{\downarrow}(\mathbb{Z}, <, =)$ and $\text{CLTL}^{\downarrow}(\mathbb{R}, <, =)$ are PSPACE-complete (Section 3.2). Flat fragments of plain LTL versions have been studied in [20,10] (see also in [21, Section 5] the design of a flat logical temporal language for model-checking pushdown machines) and our definition of flatness takes advantage in a non-trivial way of the polarity of 'until' subformulae occurring in a formula. This is a standard way to restrict the interplay between modalities and quantifiers, see e.g. [22,10,23]. Although we do not claim that flat formulae are especially interesting in practice, they cover non-trivial uses of the freeze quantifier. However, they cannot express the property that a variable at distinct points takes distinct values.

Along the paper, we consider the satisfiability problem, but as shown in Section 2.3, our results extend to the model-checking problem.

CLTL$^\downarrow$($\mathcal{D}$) extends naturally the freeze-free fragment CLTL($\mathcal{D}$), and we show that it increases strictly the expressive power (Proposition 1). However, we prove that significant fragments of CLTL$^\downarrow$($\mathcal{D}$) are as expressive as the full language, for instance by recording only values of flexible variables at the current state or by allowing only rigid variables in atomic formulae (see Section 2.4).

Apart from the technical contributions of the paper, we provide comparisons with several works which involve freeze-like operators, such as in first-order quantification, in timed LTL, in hybrid logics with reference pointers, to quote a few examples.

**Structure of the paper.** In Section 2, we present Constraint LTL with the freeze quantifier, satisfiability and model-checking problems of interest, and consider relative expressivity. Section 3 contains decidability and complexity results when the underlying concrete domain is finite or with restricting to the flat fragment. In Section 4, we show that CLTL$^\downarrow$($\mathbb{N}, =$) is $\Sigma_1^1$-complete. Related work is discussed in Section 5. In Section 6, we conclude and enumerate a few open problems.

## 2 Constraint LTL with the freeze quantifier

### 2.1 Syntax and semantics

A constraint system is a set, called the domain, with a countable family of relations on this set. Let $\mathcal{D} = (D, (R_i)_{i \in I})$ be a constraint system. We define the logic CLTL$^\downarrow$($\mathcal{D}$) by giving its syntax and semantics.

**Syntax.** Let FleVarSet and RigVarSet be countable sets of variables which are respectively called *flexible variables* and *rigid variables*. Terms are given

by the grammar:

$$t ::= \underbrace{\mathbf{X} \cdots \mathbf{X}}_{n \text{ times}} x \mid y$$

where $x$ is in FleVarSet and $y$ is in RigVarSet. We use $\mathbf{X}^n$ as an abbreviation for $\underbrace{\mathbf{X} \cdots \mathbf{X}}_{n \text{ times}}$. Formulae are given by the grammar:

$$\phi ::= R(t_1, \ldots, t_n) \mid \neg \phi \mid \phi_1 \wedge \phi_2 \mid \mathbf{X}\phi \mid \phi_1 \mathbf{U} \phi_2 \mid \downarrow_{y=\mathbf{X}^n x} \phi$$

where $R$ ranges over the predicate symbols associated to the relations in $(R_i)_{i \in I}$, $x$ over FleVarSet, and $y$ over RigVarSet. Note that we use $\mathbf{X}$ for denoting either the $n^{\text{th}}$ next value $\mathbf{X}^n x$ of the variable $x$ or the formula $\mathbf{X}\phi$. We define the Boolean constants, and the temporal operators 'sometimes' and 'always', as the following abbreviations: $\top \stackrel{\text{def}}{=} R(t_1, \ldots, t_n) \vee \neg R(t_1, \ldots, t_n)$, $\mathbf{F}\phi \stackrel{\text{def}}{=} \top \mathbf{U} \phi$, $\bot \stackrel{\text{def}}{=} R(t_1, \ldots, t_n) \wedge \neg R(t_1, \ldots, t_n)$, and $\mathbf{G}\phi \stackrel{\text{def}}{=} \neg \mathbf{F} \neg \phi$.

Let FleVars($\phi$) and RigVars($\phi$) denote the sets of all flexible and rigid (respectively) variables which occur in $\phi$.

*Freeze-free fragment.* CLTL($\mathcal{D}$) is the fragment of CLTL$^\downarrow$($\mathcal{D}$) with no rigid variables and hence without freeze quantifier.

*Flat fragment.* We say that the occurrence of a subformula in a formula is *positive* if it occurs under an even number of negations, otherwise it is *negative*. The *flat fragment of CLTL$^\downarrow$($\mathcal{D}$)* is the restriction of CLTL$^\downarrow$($\mathcal{D}$) where, for any subformula $\phi_1 \mathbf{U} \phi_2$, if it is positive then $\downarrow$ does not occur in $\phi_1$, and if it is negative then $\downarrow$ does not occur in $\phi_2$.

More precisely, the flat fragment consists of the following formulae $\varphi$. Subformulae $\varphi$ are positive, whereas subformulae $\varphi^-$ are negative.

$$\varphi ::= R(t_1, \ldots, t_n) \mid \neg \varphi^- \mid \varphi_1 \wedge \varphi_2 \mid \mathbf{X}\varphi \mid \psi \mathbf{U} \varphi \mid \downarrow_{y=\mathbf{X}^n x} \varphi$$
$$\varphi^- ::= R(t_1, \ldots, t_n) \mid \neg \varphi \mid \varphi_1^- \wedge \varphi_2^- \mid \mathbf{X}\varphi^- \mid \varphi^- \mathbf{U} \psi \mid \downarrow_{y=\mathbf{X}^n x} \varphi^-$$
$$\psi ::= R(t_1, \ldots, t_n) \mid \neg \psi \mid \psi_1 \wedge \psi_2 \mid \mathbf{X}\psi \mid \psi_1 \mathbf{U} \psi_2$$

**Semantics.** A model $\sigma : \mathbb{N} \to (\mathsf{FleVarSet} \to D)$ is a sequence of mappings from FleVarSet to $D$. For any $i \in \mathbb{N}$, we write $\sigma^i$ for the model defined by $\sigma^i(j) = \sigma(i+j)$ for every $j \geq 0$. An environment $\rho$ is a mapping from RigVarSet to $D$. We write $\rho[x \mapsto v]$ for the environment mapping $x$ to $v \in D$, and any other variable $y$ to $\rho(y)$. The semantics of terms is given by:

$$[\![\mathbf{X}^n x]\!]_{\sigma,\rho} = \sigma(n)(x) \quad \text{if } x \text{ is in FleVarSet}$$
$$[\![y]\!]_{\sigma,\rho} = \rho(y) \quad \text{if } y \text{ is in RigVarSet}$$

The semantics of formulae is given by the following satisfaction relation. (Note that we use $R$ for both a relation symbol and the relation it denotes.)

- $\sigma \models_\rho R(t_1, \ldots, t_n)$ iff $([\![t_1]\!]_{\sigma,\rho}, \ldots, [\![t_n]\!]_{\sigma,\rho}) \in R$,
- $\sigma \models_\rho \neg \phi$ iff $\sigma \not\models_\rho \phi$,
- $\sigma \models_\rho \phi_1 \wedge \phi_2$ iff $\sigma \models_\rho \phi_1$ and $\sigma \models_\rho \phi_2$,
- $\sigma \models_\rho \mathbf{X} \phi$ iff $\sigma^1 \models_\rho \phi$,
- $\sigma \models_\rho \phi_1 \mathbf{U} \phi_2$ iff there exists $i$ such that $\sigma^i \models_\rho \phi_2$ and for all $j < i$, $\sigma^j \models_\rho \phi_1$,
- $\sigma \models_\rho \downarrow_{y=\mathbf{X}^n x} \phi$ iff $\sigma \models_{\rho[y \mapsto \sigma(n)(x)]} \phi$.

2.2 *Examples*

As a first example, consider the formula

$$\phi_\infty^x \stackrel{\text{def}}{=} \mathbf{G} \downarrow_{y=x} \mathbf{XG}\, x \neq y$$

which states that the values of the variable $x$ at different points in time are mutually distinct. This is interesting for the verification of cryptographic protocols, where nonces are variables which have to be fresh, i.e. they cannot take twice the same value.

As a second example, we consider a process requesting memory blocks. Let us assume two flexible variables $o$ (for operator) and $a$ (for argument) such that $o$ takes its values in the finite domain $\{\textit{Malloc}, \textit{Access}, \textit{Free}\}$ and $a$ takes its values in an infinite set of memory locations.

We use $\textit{Malloc}(x)$, $\textit{Access}(x)$ and $\textit{Free}(x)$ as respective abbreviations for $o = \textit{Malloc} \wedge a = x$, $o = \textit{Access} \wedge a = x$, and $o = \textit{Free} \wedge a = x$ ($x$ is a rigid variable).

We can easily express the following properties in $\text{CLTL}^{\downarrow}(\mathcal{D})$.

- As soon as a memory location is freed, either it is never accessed again, or it is not accessed until it is allocated again:

$$\mathbf{G}(o = \textit{Free} \Rightarrow \downarrow_{x=a} (\mathbf{G} \neg \textit{Access}(x) \vee \neg \textit{Access}(x) \mathbf{U} \textit{Malloc}(x)))$$

- When a memory location is allocated, it will either be freed in the future or will always be eventually accessed (so that we do not waste memory):

$$\mathbf{G}(o = \mathit{Malloc} \;\Rightarrow\; \downarrow_{x=a} (\mathbf{F}\mathit{Free}(x) \;\vee\; \mathbf{GF}\mathit{Access}(x)))$$

## 2.3 Satisfiability and model-checking problems

We recall below the problems we are interested in.

*Satisfiability problem for CLTL$^\downarrow$($\mathcal{D}$):*
**instance**: a CLTL$^\downarrow$($\mathcal{D}$) formula $\phi$;
**question**: is there a model $\sigma$ and an environment $\rho$ such that $\sigma \models_\rho \phi$?

Without loss of generality we can assume that no rigid variable occurs free in $\phi$, which means that $\rho$ is not essential above.

The model-checking problem rests on $\mathcal{D}$-automata which are constraints automata. A $\mathcal{D}$-automaton is simply a Büchi automaton with alphabet a finite set of Boolean combinations of atomic CLTL$^\downarrow$($\mathcal{D}$) formulae with terms of the form $x$ and $\mathbf{X}x$ ($x \in \mathsf{FleVarSet}$). In a $\mathcal{D}$-automaton, letters on transitions induce constraints between the variables of the current state and the variables of the next state as done in [10]. Alternatively, labelling the transitions by CLTL$^\downarrow$($\mathcal{D}$) formulae (as done in [24]) would not modify essentially the decidability status of model-checking problems considered in this paper.

*Model-checking problem for CLTL$^\downarrow$($\mathcal{D}$):*
**instance**: a $\mathcal{D}$-automaton $\mathcal{A}$ and a CLTL$^\downarrow$($\mathcal{D}$) formula $\phi$;
**question**: are there a symbolic $\omega$-word $v = \phi_0, \phi_1, \ldots$ accepted by $\mathcal{A}$, a model $\sigma$ (a realisation of $v$) and an environment $\rho$ such that $\sigma \models_\rho \phi$ and for every $i \geq 0$, $\sigma^i \models_\rho \phi_i$?

It is not difficult to show that as soon as $\mathcal{D}$ is non-trivial the satisfiability problem and the model-checking problem are reducible to each other in logarithmic space following techniques from [19]. In the sequel, we prove results for the satisfiability problem but one has to keep in mind that our results extend to the model-checking problem.

## 2.4 Expressive power

**The freeze quantifier strictly increases expressive power.** In order to show formally that the freeze quantifier is powerful, we show that CLTL$^\downarrow$($\mathbb{N}, =$) is strictly more expressive than its freeze-free fragment CLTL($\mathbb{N}, =$). In fact, $\phi_\infty^x$ is an example of a formula $\phi$ in CLTL$^\downarrow$($\mathbb{N}$, =) with no free rigid variable for which there is no equivalent formula $\psi$ in CLTL($\mathbb{N}$, =). The result will follow from the following property.

**Lemma 1** *Every satisfiable formula $\phi$ in CLTL($\mathbb{N}$, =) has a model which contains only finitely many distinct values. Moreover, the number of distinct values is polynomial in $|\phi|$.*

*Proof.* Let $\phi$ be a formula in CLTL($\mathbb{N}$, =) with variables in $\{x_1, \ldots, x_n\}$ and $k$ be equal to 1 plus the maximal $j$ such that $\mathbf{X}^j x_i$ occurs in $\phi$ for some flexible variable $x_i$. Let $C$ be the finite set of constraints of the form $\mathbf{X}^{j_1} x_{i_1} = \mathbf{X}^{j_2} x_{i_2}$ with $0 \leq j_1, j_2 \leq k-1$ and $i_1, i_2 \in \{1, \ldots, n\}$.

We define a total ordering on $\{1, \ldots, n\} \times \mathbb{N}$ as follows: $\langle i, j \rangle < \langle i', j' \rangle$ iff $j < j'$ or ($j = j'$ and $i < i'$). Given a model $\sigma : \mathbb{N} \to (\mathsf{FleVarSet} \to \mathbb{N})$, we build a model $\sigma' : \mathbb{N} \to (\mathsf{FleVarSet} \to \{1, \ldots, k \times n\})$ such that $\sigma \models \phi$ iff $\sigma' \models \phi$.

If $x$ is a flexible variable not occurring in $\phi$, $\sigma'(i)(x) = 1$ for every $i \geq 0$. Otherwise $\sigma'(0)(x_1) = 1$ ($\langle 1, 0 \rangle$ is minimal wrt $<$). Now suppose that for every $\langle i', j' \rangle < \langle i, j \rangle$, $\sigma'(j')(x_{i'})$ has been already defined. We shall define $\sigma'(j)(x_i)$. If for some $\langle i', j' \rangle$ in $\{\langle i'', j'' \rangle : 0 \leq j - j'' \leq k-1,\ 1 \leq i'' \leq n,\ \langle i'', j'' \rangle < \langle i, j \rangle\}$, $\sigma(j')(x_{i'}) = \sigma(j)(x_i)$ then $\sigma'(j)(x_i)$ takes the value $\sigma'(j')(x_{i'})$. Otherwise, $\sigma'(j)(x_i)$ takes an arbitrary value from the set

$$\{1, \ldots, k \times n\} \setminus \{\sigma(j'')(x_{i''}) : 0 \leq j - j'' \leq k-1,\ 1 \leq i'' \leq n, \langle i'', j'' \rangle < \langle i, j \rangle\}$$

which is always possible since the second set has strictly less that $k \times n$ elements. One can show that for all $c \in C$ and $i \geq 0$, $\sigma'^i \models c$ iff $\sigma^i \models c$. Hence, $\sigma \models \phi$ iff $\sigma' \models \phi$. □

**Proposition 1** *No formula of CLTL($\mathbb{N}$, =) is equivalent to the formula $\phi_\infty^x$ of CLTL$^\downarrow$($\mathbb{N}$, =).*

The flatness concept is only related to occurrences of the freeze quantifier and for instance the formulae of the form $\phi_\infty^x$ do not belong to the flat fragment. By contrast, $\neg \phi_\infty^x$ belongs to the flat fragment of CLTL$^\downarrow$($\mathbb{N}$, =). By Proposition 1, the flat fragment of CLTL$^\downarrow$($\mathbb{N}$, =) is therefore strictly more expressive than CLTL($\mathbb{N}$, =) since CLTL($\mathbb{N}$, =) is closed under negation.

**Equivalent syntactic restrictions.** We now show that expressiveness of CLTL$^\downarrow$($\mathcal{D}$) does not change if we restrict the freeze quantifier to refer only to flexible variables in the current state, or if we restrict atomic formulae to contain only rigid variables, or with both restrictions. Therefore, those restrictions could have been incorporated into the definition of the logic. However, we chose to allow terms of the form $\mathbf{X}^n x$ with flexible $x$ in atomic formulae in order to have CLTL($\mathcal{D}$) as the freeze-free fragment, and to allow the freeze quantifier to refer to the future so that formulae would be closed under substitution of terms.

**Proposition 2** *For any formula $\phi$ of $CLTL^{\downarrow}(\mathcal{D})$, there exists an equivalent formula $\phi'$ such that:*

**(I)** *any occurence of $\downarrow$ in $\phi'$ is of the form $\downarrow_{y=x}$;*
**(II)** $\mathsf{FleVars}(\phi') = \mathsf{FleVars}(\phi)$;
**(III)** $\mathsf{RigVars}(\phi') = \mathsf{RigVars}(\phi)$.

*Proof.* By structural induction on $\phi$, it suffices to prove the statement for formulae of the form $\downarrow_{y=\mathbf{X}^n x} \phi'$ where $\phi'$ satisfies (I).

This can be done by induction on $n$. The base case $n = 0$ is trivial. For the inductive step, we use structural induction on $\phi'$. The most difficult case is $\phi' = \phi'_1 \mathbf{U} \phi'_2$. We then have

$$\downarrow_{y=\mathbf{X}^{n+1}x} \phi'$$
$$\equiv \downarrow_{y=\mathbf{X}^{n+1}x} \phi'_2 \vee (\phi'_1 \wedge \mathbf{X}\phi')$$
$$\equiv (\downarrow_{y=\mathbf{X}^{n+1}x} \phi'_2) \vee ((\downarrow_{y=\mathbf{X}^{n+1}x} \phi'_1) \wedge \mathbf{X} \downarrow_{y=\mathbf{X}^n x} \phi')$$

and the induction hypotheses apply to each of the three freeze subformulae. □

It is worth observing that in the worst case, in the proof of Proposition 2, $\phi'$ can be exponentially larger than $\phi$.

**Proposition 3** *For any formula $\phi$ of $CLTL^{\downarrow}(\mathcal{D})$, there exists an equivalent formula $\phi'$ such that:*

- *atomic formulae in $\phi'$ contain only rigid variables;*
- *if any occurence of $\downarrow$ in $\phi$ is of the form $\downarrow_{y=x}$, then the same is true of $\phi'$;*
- $\mathsf{FleVars}(\phi') = \mathsf{FleVars}(\phi)$;
- $|\mathsf{RigVars}(\phi')| = \max\{|\mathsf{RigVars}(\phi)|, k\}$, *where $k$ is the maximum number of distinct terms in any atomic subformula of $\phi$.*

*Proof.* $\phi'$ is constructed from $\phi$ by translating only atomic subformulae of $\phi$. For example, $R(\mathbf{X}^2 x_1, y_1, \mathbf{X}^3 x_2, \mathbf{X}^2 x_3, x_4, y_2, x_4)$, where $x_i \in \mathsf{FleVarSet}$ and $y_i \in \mathsf{RigVarSet}$, is translated to

$$\downarrow_{y_3=x_4} \mathbf{X}^2 \downarrow_{y_4=x_1}\downarrow_{y_5=x_3} \mathbf{X}^1 \downarrow_{y_6=x_2} R(y_4, y_1, y_6, y_5, y_3, y_2, y_3)$$

where $y_3$, ..., $y_6$ are drawn from $\mathsf{RigVars}(\phi) \setminus \{y_1, y_2\}$. If that set does not have enough elements, new rigid variable names are used. The latter can then be reused in translations of other atomic subformulae. □

**Flexible and finitary variables.** If the domain $D$ has at least two elements, and if the equality predicate is present, then formulae and models of $\text{CLTL}^{\downarrow}(\mathcal{D})$ with $n \geq 2$ flexible variables can be translated to the fragment with only one flexible variable.

**Proposition 4** *Let $\mathcal{D}$ be a constraint system with at least two elements and equality. For any formula $\phi$ of $\text{CLTL}^{\downarrow}(\mathcal{D})$, one can compute in logarithmic space a formula $\phi'$ of $\text{CLTL}^{\downarrow}(\mathcal{D})$ with a unique flexible variable and the same set of rigid variables as $\phi$, such that $\phi$ is satisfiable iff $\phi'$ is satisfiable.*

*Proof.* Let $\phi$ be a formula of $\text{CLTL}^{\downarrow}(\mathcal{D})$ with flexible variables $x_1$, ..., $x_n$. We shall build in logspace a formula $\phi'$ of $\text{CLTL}^{\downarrow}(\mathcal{D})$ with only one flexible variable $x'$ and the same set of rigid variables as $\phi$, such that $\sigma' \models_\rho \phi'$ iff there exists $\sigma$ with $\sigma \models_\rho \phi$ and $\sigma'$ is an encoding of $\sigma$ in the following sense. A valuation $\sigma(i) : \{x_1, \ldots, x_n\} \to D$ is encoded by $2n + 4$ consecutive values of $x'$ in $\sigma'$ which form a sequence

$$d_1^i, d_0^i, d_0^i, d_0^i, d_1^i, \sigma(i)(x_1), d_2^i, \sigma(i)(x_2), \ldots, d_n^i, \sigma(i)(x_n)$$

Using the equality predicate, the values $d_j^i$ are constrained in $\phi'$ so that three consecutive equal values occur in $\sigma'$ only at the beginnings of sequences which encode valuations in $\sigma$.

The formula $\phi'$ is a conjunction $\phi^{\text{enc}} \wedge T(\phi)$ where $\phi^{\text{enc}}$ enforces that models are sequences of length $2n+4$ of the above form (details are omitted here). Formula $T(\phi)$ is inductively defined as follows where $\texttt{start} = \mathbf{X}(x' = \mathbf{X}x' \wedge x' = \mathbf{XX}x')$:

- $T(R(t_1, \ldots, t_m)) = R(T(t_1), \ldots, T(t_m))$ where $T(y) = y$ if $y$ is rigid and $T(\mathbf{X}^k x_i) = \mathbf{X}^{k \times (2n+4)+3+2i} x'$,
- $T$ is homomorphic for Boolean connectives,
- $T(\downarrow_{y=\mathbf{X}^k x_i} \phi_1) = \downarrow_{y=T(\mathbf{X}^k x_i)} T(\phi_1)$,
- $T(\phi_1 \mathbf{U} \phi_2) = (\texttt{start} \Rightarrow T(\phi_1)) \mathbf{U}(\texttt{start} \wedge T(\phi_2))$,
- $T(\mathbf{X}\phi_1) = \mathbf{X}^{2n+4} T(\phi_1)$. □

The logics $\text{CLTL}^{\downarrow}(\mathcal{D})$ as defined in Section 2.1 do not in general have propositional variables. If $D$ has at least two elements and equality, then propositional flexible variables, or a flexible variable ranging over a finite alphabet, can be encoded using additional flexible variables over $D$ and equality. A translation as above can then be employed to reduce the number of flexible variables.

For ease of expression, to avoid unnecessary constructs, and because equality on the domain is not necessarily present, arbitrarily many flexible variables and no special finitary variables are considered in the rest of the paper.

## 3 Decidability results

*3.1 Finite domain case*

In this section, we basically show that, when $\mathcal{D}$ is finite (with at least two elements) and contains the equality predicate, $\text{CLTL}^{\downarrow}(\mathcal{D})$ is ExpSpace-complete. In Theorem 1 below, we establish that ExpSpace-hardness is very common when the freeze quantifier is present.

**Theorem 1** *Let $\mathcal{D}$ be a constraint system with equality such that the underlying domain $D$ contains at least two elements. The satisfiability problem for $CLTL^{\downarrow}(\mathcal{D})$ is ExpSpace-hard.*

*Proof.* We prove this result by a reduction from an ExpSpace-complete tiling problem (see e.g. [25]). A tile is a unit square of one of several types and the tiling problem we consider is specified by means of a finite set $T$ of tile types (say $T = \{t_1, \ldots, t_l\}$), two binary relations $H$ (horizontal matching relation) and $V$ (vertical matching relation) over $T$ and two distinguished tile types $t_{init}$, $t_{final} \in T$. The problem consists in determining whether, for a given number $n$ in unary, the region $[0, \ldots, 2^n - 1] \times [0, \ldots, k - 1]$ of the integer plane for some $k$ can be tiled consistently with $H$ and $V$, $t_{init}$ is the left bottom tile, and $t_{final}$ is the right upper tile.

Given an instance $I = \langle T, t_{init}, t_{final}, n \rangle$ of the tiling problem, we build a $\text{CLTL}^{\downarrow}(\mathcal{D})$ formula $\phi_I$ such that $I = \langle T, t_{init}, t_{final}, n \rangle$ has a solution iff $\phi_I$ is $\text{CLTL}^{\downarrow}(\mathcal{D})$ satisfiable.

We consider the following flexible variables:

- $c_1, \ldots, c_n$ are variables that allow to count until $2^n$ and $x_0, x_1$ are variables that will play the role of 0 and 1, respectively; there are corresponding rigid variables $c'_1, \ldots, c'_n$; each element $\langle \alpha, i \rangle$ of a row $[0, \ldots, 2^n - 1] \times \{i\}$ such that the binary representation of $\alpha$ is $b_1 \ldots b_n$, satisfies $c_j = x_0$ iff $b_j = 0$ for every $j \in \{1, \ldots, n\}$;
- for $t \in T$, $z_t^1, z_t^2$ are variables such that $D_t := z_t^1 = z_t^2$ is the formula encoding the fact that at a certain position of the integer plane the tile $t$ is present. There are also rigid variables $z_t^{1'}, z_t^{2'}$ and $D'_t := z_t^{1'} = z_t^{2'}$;
- $end_1, end_2$ such that $\text{END} := end_1 = end_2$;

The formula $\phi_I$ is the conjunction of the following formulae:

- The region of the integer plane for the solution is finite:
$$\neg \text{END} \land (\neg \text{END} \mathbf{U}(c_1 = \cdots = c_n = x_0 \land \mathbf{G}\ \text{END}))$$

- $x_0$ and $x_1$ behave as different constants:
$$\neg(x_0 = x_1) \land \mathbf{G}(x_0 = \mathbf{X}x_0 \land x_1 = \mathbf{X}x_1)$$

- There is exactly one tile per element of the plane region:
$$\mathbf{G}(\neg \text{END} \Rightarrow \bigvee_{t \in T}(D_t \land \bigwedge_{t' \neq t} \neg D_{t'}))$$

- Constraint on the right upper tile:
$$\mathbf{F}(\bigwedge_{1 \leq i \leq n}(c_i = x_1) \land \neg \text{END} \land D_{t_{final}} \land \mathbf{X}\text{END})$$

- Constraint on the left bottom tile:
$$\bigwedge_{1 \leq i \leq n}(c_i = x_0) \land D_{t_{init}}$$

- Incrementation of the counters $c_1, \ldots, c_n$:
$$\mathbf{G}(\bigvee_{2 \leq i \leq n+1}((\bigwedge_{i \leq j \leq n} c_j = x_1) \land c_{i-1} = x_0 \land \neg \text{END})$$
$$\Rightarrow (\bigwedge_{1 \leq j \leq i-2}(c_j = \mathbf{X}c_j) \land \mathbf{X}c_{i-1} = x_1 \land \bigwedge_{i \leq j \leq n}(\mathbf{X}c_j = x_0))))$$

- Limit condition for the incrementation of the counters $c_1, \ldots, c_n$:
$$\mathbf{G}((\neg \mathbf{X}\text{END} \land c_1 = \cdots = c_n = x_1) \Rightarrow \mathbf{X}(c_1 = \cdots = c_n = x_0))$$

- Horizontal consistency:
$$\mathbf{G}(\overbrace{\neg(c_1 = \cdots = c_n = x_1)}^{\text{not the last element of a row}} \land \neg \text{END} \Rightarrow \bigwedge_{t \in T}(D_t \Rightarrow \bigvee_{\langle t,t' \rangle \in H} \mathbf{X}D_{t'}))$$

- Vertical consistency:
$$\mathbf{G}(\neg \text{END} \land \overbrace{\mathbf{F}(\mathbf{X}\neg \text{END} \land c_1 = \ldots = c_n = x_1)}^{\text{not on the last row}} \Rightarrow$$
$$\downarrow_{c'_1=c_1} \cdots \downarrow_{c'_n=c_n} \downarrow_{z^{1'}_{t_1}=z^1_{t_1}} \downarrow_{z^{2'}_{t_1}=z^2_{t_1}} \cdots \downarrow_{z^{1'}_{t_k}=z^1_{t_k}} \downarrow_{z^{2'}_{t_k}=z^2_{t_k}}$$
$$\mathbf{X}((\neg \bigwedge_{1 \leq i \leq n} c'_i = c_i)\mathbf{U}(\bigwedge_{1 \leq i \leq n} c'_i = c_i \land \bigwedge_{t \in T}(D'_t \Rightarrow \bigvee_{\langle t,t' \rangle \in V} \mathbf{X}D_{t'})))$$

It is not difficult to show that the instance $I = \langle T, t_{init}, t_{final}, n \rangle$ has a solution iff $\phi_I$ is CLTL$^\downarrow(\mathcal{D})$ satisfiable. □

This is reminiscent to the ExpSpace-hardness of Timed Propositional Temporal Logic (TPTL) [12, Theorem 2], PLTL+Now (NLTL) [26, Proposition 4.7] and a variant of the guarded fragment with transitivity [27, Theorem 2]. Our ExpSpace-hardness proof is in the same vein since basically in CLTL$^\downarrow(\mathcal{D})$ we are able to count till $2^n$ using only a number of resources polynomial in $n$ and we can compare the truth value of atomic formulae in states of "temporal distance" exactly $2^n$.

Our proof is a slight variant of the proof of [14, Theorem 6]: instead of using integer periodicity constraints to count till $2^n$, $n$ binary counters are used. Observe also that the resulting formula is not flat because of the encoding of vertical consistency.

If we replace **U** by **F**, then NExpTime-hardness can be shown by reducing from the $n \times n$ tiling problem with $n$ encoded in binary.

Finiteness of $\mathcal{D}$ allows us to show the decidability of CLTL$^\downarrow(\mathcal{D})$.

**Theorem 2** *Let $\mathcal{D}$ be a finite constraint system. The satisfiability problem for CLTL$^\downarrow(\mathcal{D})$ is in ExpSpace.*

*Proof.* Assume that $D = \{d_1, \ldots, d_l\}$. We introduce an auxiliary constraint system $\mathcal{D}' = \langle D, P_1, \ldots, P_l \rangle$ such that $P_i = \{d_i\}$. For convenience, we write $x = d_i$ instead of $P_i(x)$. We shall show how to reduce the satisfiability problem for CLTL$^\downarrow(\mathcal{D})$ into the satisfiability problem for CLTL($\mathcal{D}'$). PSpace-membership of CLTL($\mathcal{D}'$) is not very difficult to show and it is a direct consequence of [14, Theorem 4].

We introduce a translation T from CLTL$^\downarrow(\mathcal{D})$ formulae into CLTL($\mathcal{D}'$) formulae defined as follows:

- T is homomorphic for the Boolean operators and the temporal operators,
- $\mathrm{T}(\mathsf{R}(\alpha_1, \ldots, \alpha_n)) \stackrel{\text{def}}{=} (\bigvee_{R(d_{i_1}, \ldots, d_{i_n})} (\alpha_1 = d_{i_1} \wedge \cdots \wedge \alpha_n = d_{i_n}))$.

So far, the translation can be done in polynomial time and logarithmic space since $|D|^m$ is a constant of CLTL$^\downarrow(\mathcal{D})$ where $m$ is the maximal arity of relations in $\mathcal{D}$. The last clause of T is related to the freeze quantifier:

$$\mathrm{T}(\downarrow_{x'=\alpha} \psi) \stackrel{\text{def}}{=} \bigwedge_{d_i \in D} (\alpha = d_i) \Rightarrow \mathrm{T}(\psi)^{x'=d_i},$$

where $\mathrm{T}(\psi)^{x'=d_i}$ is obtained from $\mathrm{T}(\psi)$ by replacing every occurrence of $x' = d_j$ with $j \neq i$ by $\bot$ and every occurrence of $x' = d_i$ by $\top$. This step requires an exponential blow up and therefore $|\mathrm{T}(\phi)|$ is exponential in $|\phi|$. It is easy to show that $\phi$ is CLTL$^\downarrow$($\mathcal{D}$) satisfiable iff T($\phi$) is CLTL($\mathcal{D}'$) satisfiable. Since T may cause at most an exponential blow up and CLTL($\mathcal{D}'$) is in PSPACE, we obtain that CLTL$^\downarrow$($\mathcal{D}$) satisfiability is in EXPSPACE. □

Our proof can be easily adapted if the freeze quantifier is replaced by the full existential quantifier ∃.

**Corollary 1** *Let $\mathcal{D}$ be a finite constraint system with equality such that the underlying domain D contains at least two elements. The satisfiability problem for CLTL$^\downarrow$($\mathcal{D}$) is* EXPSPACE*-complete.*

A formula $\phi \in$ CLTL$^\downarrow$($\mathcal{D}$) is of $\downarrow$-height $k$, for some $k \geq 0$, whenever every branch of the formula tree of $\phi$ has at most $k$ freeze quantifiers. For example, the formula $\downarrow_{x'=x} (y = x')\mathbf{U} \downarrow_{x'=z} y = x'$ is of $\downarrow$-height 2.

**Corollary 2** *Let $\mathcal{D}$ be a finite constraint system. For every $k \geq 0$, the satisfiability problem for CLTL$^\downarrow$($\mathcal{D}$) restricted to formulae of $\downarrow$-height $k$ is in* PSPACE.

The complexity of CLTL$^\downarrow$($\mathcal{D}$) with finite $\mathcal{D}$ and restricted to the 'sometimes' operator **F** is still open. (NEXPTIME-hardness and EXPSPACE upper bound are known.)

### 3.2 Flat fragment between CLTL($\mathcal{D}$) and CLTL$^\downarrow$($\mathcal{D}$)

The main result of this section is to show that the freeze quantifier in the flat fragment of CLTL$^\downarrow$($\mathcal{D}$) can be encoded faithfully into CLTL($\mathcal{D}$) even though flat CLTL$^\downarrow$($\mathcal{D}$) can be more expressive than CLTL($\mathcal{D}$), see for instance the case with $\mathcal{D} = \langle \mathbb{N}, = \rangle$ in Section 2.4. However, as shown below, satisfiability for flat CLTL$^\downarrow$($\mathbb{N}, =$) can be reduced in logarithmic space to satisfiability for CLTL($\mathbb{N}, =$). By analogy, CTL* model-checking can be reduced to LTL model-checking [28] even though CTL* is more expressive than LTL.

It is worth observing that our concept of flatness restricts the interplay between future-time operators and the freeze quantifier as done in [22,10,23] to limit the interaction between modalities and freeze-like quantifiers. In order to understand why flat formulae are more manageable, in a formula like $\downarrow_{y=x} \mathbf{F}\phi$ that is flat, only the current value of $x$ needs to be stored. By contrast, in a formula like $\mathbf{G} \downarrow_{y=x} \phi$ that is not flat, one needs to store as many values of $x$ as there are positions.

We assume that the flexible variables of CLTL$^\downarrow$($\mathcal{D}$) are $\{x_0, x_1, \ldots\}$ and the rigid variables of CLTL$^\downarrow$($\mathcal{D}$) are $\{y_0, y_1, \ldots\}$. For ease of presentation, we assume that the flexible variables of CLTL($\mathcal{D}$) are composed of the following two disjoint sets: $\{x_0, x_1, \ldots\}$ and $\{y_0^{\text{new}}, y_1^{\text{new}}, \ldots\}$. We define a map $u$ from the flat fragment $\text{CLTL}^{\downarrow}(\mathcal{D})$ into $\text{CLTL}(\mathcal{D})$ as follows: $u$ replaces each $y_j$ by $y_j^{\text{new}}$ in atomic formulae, it is homomorphic for Boolean and temporal operators, and

$$u(\downarrow_{y=\mathbf{X}^n x} \psi) \stackrel{\text{def}}{=} y^{\text{new}} = \mathbf{X}^n x \;\wedge\; \mathbf{G}(y^{\text{new}} = \mathbf{X} y^{\text{new}}) \;\wedge\; u(\psi)$$

It is easy to show that $u(\phi)$ can be computed in logarithmic space in $|\phi|$.

**Proposition 5** *Let $\mathcal{D}$ be a constraint system with equality. For any formula $\phi$ of the flat fragment of $\text{CLTL}^{\downarrow}(\mathcal{D})$, $\phi$ is $\text{CLTL}^{\downarrow}(\mathcal{D})$ satisfiable iff $u(\phi)$ is $\text{CLTL}(\mathcal{D})$ satisfiable.*

*Proof.* Given a model $\sigma$ of $\text{CLTL}^{\downarrow}(\mathcal{D})$, an environment $\rho$ and a formula $\phi$ we say that the model $\sigma'$ of $\text{CLTL}(\mathcal{D})$ agrees with $\sigma$, $\rho$ and $\phi$ iff for all $i, j \geq 0$, $\sigma(i)(x_j) = \sigma'(i)(x_j)$ and for all free rigid variable $y_j$ in $\phi$ and $i \geq 0$, $\sigma'(i)(y_j^{\text{new}}) = \rho(y_j)$.

We shall use the following basic properties:

- $u(\psi) = \psi$ if $\psi$ belongs to $\text{CLTL}(\mathcal{D})$.
- If $\sigma'$ agrees with $\sigma$, $\rho$ and $\psi$ then $(\sigma')^i$ agrees with $\sigma^i$, $\rho$ and $\psi$ for every $i \geq 0$.

Given the occurrence of a subformula $\psi$ in $\phi$ with positive [resp. negative] polarity, we write the sign $s_\psi$ to denote the empty string [resp. $\neg$]. By abusing notation, we do not distinguish subformulae from occurrences.

We shall show by structural induction that for any occurrence of a subformula $\psi$ in $\phi$, for all models $\sigma$ of $\text{CLTL}^{\downarrow}(\mathcal{D})$ and environment $\rho$, $\sigma \models_\rho s_\psi \psi$ iff there is $\sigma'$ that agrees with $\sigma$, $\rho$ and $\psi$ such that $\sigma' \models s_\psi u(\psi)$. Statement of the lemma is then immediate.

The base case with atomic formulae and the cases in the induction step with $\neg$, $\wedge$ and $\mathbf{X}$ are by an easy verification. By way of example, we treat the case with $\psi = \neg \psi'$ with negative polarity. So $\psi'$ occurs with positive polarity. Let $\sigma$ be a model and $\rho$ be an environment such that $\sigma \models_\rho \neg\neg\psi'$. The statements below are equivalent:

- $\sigma \models_\rho \neg\neg\psi'$,
- $\sigma \models_\rho \psi'$,
- there is $\sigma'$ that agrees with $\sigma$, $\rho$ and $\psi'$ such that $\sigma' \models u(\psi')$ (by (IH) and change of polarity),
- there is $\sigma'$ that agrees with $\sigma$, $\rho$ and $\psi'$ such that $\sigma' \models \neg u(\neg \psi')$ (by definition of $u$).

Let us treat the remaining cases.

*Case 1*: $\psi = \psi_1 \mathbf{U} \psi_2$ with positive polarity.
Since $\phi$ belongs to the flat fragment, we have $\psi_1 = u(\psi_1)$. Let $\sigma$ be a model and $\rho$ be an environment such that $\sigma \models_\rho \psi$. The statements below are equivalent:

- $\sigma \models_\rho \psi$,
- there is $i \geq 0$ such that $\sigma^i \models_\rho \psi_2$ and for every $j < i$, $\sigma^j \models_\rho \psi_1$,
- there is $\sigma'$ that agrees with $\sigma$, $\rho$ and $\psi_2$ such that $(\sigma')^i \models u(\psi_2)$ and for every $j < i$, $(\sigma')^j \models u(\psi_1)$ (by (IH), $\psi_1 = u(\psi_1)$ and, $\sigma$ and $\sigma'$ agree on flexible variables of $\psi_1$),
- there is $\sigma'$ that agrees with $\sigma$, $\rho$ and $\psi$ such that $\sigma' \models u(\psi_1)\mathbf{U}u(\psi_2)$ ($\psi_1$ has no free rigid variable).

*Case 2*: $\psi = \psi_1 \mathbf{U} \psi_2$ with negative polarity.
Since $\phi$ belongs to the flat fragment, we have $\psi_2 = u(\psi_2)$ and both $\psi_1$ and $\psi_2$ have negative polarity. Let $\sigma$ be a model and $\rho$ be an environment such that $\sigma \models_\rho \psi$. The statements below are equivalent:

- $\sigma \models_\rho \neg\psi$,
- either there is $j \geq 0$ such that $\sigma^j \models_\rho \neg\psi_1$ and for every $j \leq i$, $\sigma^i \models_\rho \neg\psi_2$ or for every $i \geq 0$, $\sigma^i \models_\rho \neg\psi_2$,
- either there is $\sigma'$ that agrees with $\sigma$, $\rho$ and $\psi_1$ such that there is $j \geq 0$ such that $(\sigma')^j \models \neg u(\psi_1)$ and for every $j \leq i$, $(\sigma')^i \models \neg u(\psi_2)$ (by (IH) and $\psi_2 = u(\psi_2)$) or there is $\sigma'$ that agrees with $\sigma$, $\rho$ and $\psi_2$ such that for every $i \geq 0$, $(\sigma')^i \models \neg u(\psi_2)$ (by (IH)),
- there is $\sigma'$ that agrees with $\sigma$, $\rho$ and $\psi_1 \mathbf{U} \psi_2$ such that either there is $j \geq 0$ such that $(\sigma')^j \models \neg u(\psi_1)$ and for every $j \leq i$, $(\sigma')^i \models \neg u(\psi_2)$ or for every $i \geq 0$, $(\sigma')^i \models \neg u(\psi_2)$ ($\psi_2$ has no free rigid variables),
- there is $\sigma'$ that agrees with $\sigma$, $\rho$ and $\psi_1 \mathbf{U} \psi_2$ such that $\sigma' \models \neg(u(\psi_1)\mathbf{U}u(\psi_2))$.

*Case 3*: $\psi = \downarrow_{y=\mathbf{X}^n x} \psi'$.
Let $\sigma$ be a model and $\rho$ be an environment for $s_\psi$ and $\psi$. The statements below are equivalent:

- $\sigma \models_\rho s_\psi \psi$,
- $\sigma \models_{\rho[y \mapsto \sigma(n)(x)]} s_\psi \psi'$,
- there is $\sigma'$ that agrees with $\sigma$, $\rho[y \mapsto \sigma(n)(x)]$ and $\psi'$ such that $\sigma' \models s_\psi u(\psi')$ (by (IH)),
- there is $\sigma'$ that agrees with $\sigma$, $\rho[y \mapsto \sigma(n)(x)]$ and $\psi'$ such that $\sigma' \models s_\psi u(\psi')$ and $\sigma' \models \mathbf{G}(y^{\text{new}} = \mathbf{X}y^{\text{new}}) \wedge y^{\text{new}} = \mathbf{X}^n x$ ($y$ free in $\psi'$).
- there is $\sigma'$ that agrees with $\sigma$, $\rho$ and $\psi$ such that $\sigma' \models s_\psi u(\psi') \wedge \mathbf{G}(y^{\text{new}} = \mathbf{X}y^{\text{new}}) \wedge y^{\text{new}} = \mathbf{X}^n x$ ($\psi$ has less free rigid variable than $\psi'$). □

**Corollary 3** *For every constraint system $\mathcal{D}$ which contains equality, decidability of CLTL($\mathcal{D}$) implies decidability of the flat fragment of CLTL$^\downarrow$($\mathcal{D}$).*

Since CLTL($\langle \mathbb{Z}, <, = \rangle$), CLTL($\langle \mathbb{N}, <, = \rangle$) and CLTL($\langle \mathbb{R}, <, = \rangle$) are PSPACE-complete [11], we can establish the following corollary.

**Corollary 4** *Flat fragments of each of $CLTL^{\downarrow}(\langle \mathbb{Z}, <, = \rangle)$, $CLTL^{\downarrow}(\langle \mathbb{N}, <, = \rangle)$, $CLTL^{\downarrow}(\langle \mathbb{R}, <, = \rangle)$, and $CLTL^{\downarrow}(\mathcal{D})$ with $\mathcal{D}$ finite are* PSpace-*complete.*

Corollary 4 can be also adapted to the PSpace-complete constrained version of LTL introduced in [29].

## 4 Undecidability results

In this section, we shall prove that, if the domain is infinite, and if we do not restrict to flat formulae, the satisfiability problem for $\text{CLTL}^{\downarrow}(\mathcal{D})$ is undecidable even if we only have the equality predicate. More precisely, Theorem 3 below is a stronger result, stating that satisfiability is $\Sigma_1^1$-hard, even restricted to formulae with 1 flexible variable and at most 2 rigid variables. (An exposition of the analytical hierarchy can be found in [30].) A corollary of $\Sigma_1^1$-hardness is that the logic cannot be recursively axiomatised.

The following proposition complements the main result in this section, and states that, for countable and computable constraint systems $\mathcal{D}$, satisfiability for $\text{CLTL}^{\downarrow}(\mathcal{D})$ is in $\Sigma_1^1$. Hence, for a countably infinite domain, the problem in Theorem 3 is $\Sigma_1^1$-complete.

**Proposition 6** *If $D$ is countable, and $(R_i)_{i \in I}$ is a countable family of computable relations on $D$, then the satisfiability problem for $CLTL^{\downarrow}(D, (R_i)_{i \in I})$ is in $\Sigma_1^1$.*

*Proof.* Let $\phi$ be a formula of $\text{CLTL}^{\downarrow}(D, (R_i)_{i \in I})$. We can assume FleVarSet = FleVars($\phi$) and RigVarSet = RigVars($\phi$). Let $n = |\text{FleVarSet}|$, $m = |\text{RigVarSet}|$. Any model $\sigma : \mathbb{N} \to (\text{FleVarSet} \to D)$ can be encoded by functions $f_1, \ldots, f_n : \mathbb{N} \to \mathbb{N}$, and any environment $\rho : \text{RigVarSet} \to D$ as an $m$-tuple $a_1, \ldots, a_m : \mathbb{N}$. A first-order predicate on $f_1, \ldots, f_n$ and $a_1, \ldots, a_m$ which expresses that $\sigma \models_\rho \phi$ is routine to construct by structural recursion on $\phi$. We conclude that satisfiability of $\phi$ can be expressed by a $\Sigma_1^1$-sentence. □

We shall prove that the satisfiability problem for a fragment of $\text{CLTL}^{\downarrow}(D, =)$ is $\Sigma_1^1$-hard by reducing from the Recurrence Problem for nondeterministic 2-counter machines, which was shown to be $\Sigma_1^1$-hard in [12, Section 4.1].

A *nondeterministic* 2-*counter machine* $M$ consists of two counters $C_1$ and $C_2$, and a sequence of $n \geq 1$ instructions, each of which may increment or decrement one of the counters, or jump conditionally upon of the counters being zero. After the execution of a non-jump instruction, $M$ proceeds nondeterministically to one of two specified instructions. Therefore, the $l^{\text{th}}$ instruction is written as one of the following:

$$l : C_i := C_i + 1; \text{ goto } l' \text{ or goto } l''$$

$$l : C_i := C_i - 1; \text{ goto } l' \text{ or goto } l''$$

$$l : \text{if } C_i = 0 \text{ then goto } l' \text{ else goto } l''$$

We represent the configurations of $M$ by triples $\langle l, c_1, c_2 \rangle$, where $1 \leq l \leq n$, $c_1 \geq 0$, and $c_2 \geq 0$ are the current values of the location counter and the two counters $C_1$ and $C_2$, respectively. The consecution relation on configurations is defined in the obvious way, where decrementing 0 yields 0. A *computation* of $M$ is an $\omega$-sequence of related configurations, starting with the initial configuration $\langle 1, 0, 0 \rangle$. The computation is *recurring* if it contains infinitely many configurations with the value of the location counter being 1.

The Recurrence Problem is to decide, given a nondeterministic 2-counter machine $M$, whether $M$ has a recurring computation. This problem is $\Sigma_1^1$-hard.

**Theorem 3** *If $D$ is infinite, then the satisfiability problem for $CLTL^{\downarrow}(D, =)$ with $|\mathsf{FleVarSet}| = 1$ and $|\mathsf{RigVarSet}| = 2$ is $\Sigma_1^1$-hard.*

*Proof.* Suppose $M$ is a nondeterministic 2-counter machine. We construct a formula $\phi_M$ of $\mathrm{CLTL}^{\downarrow}(D,=)$ such that $|\mathsf{FleVars}(\phi)| = 1$, $|\mathsf{RigVars}(\phi)| = 2$, and $\phi_M$ is satisfiable iff $M$ has a recurring computation. The basis of the construction is an encoding of computations of nondeterministic 2-counter machines by models of $\mathrm{CLTL}^{\downarrow}(D,=)$ with one flexible variable, i.e. by $\omega$-sequences of elements of $D$. As in the proofs of [12, Theorems 6 and 7], which show $\Sigma_1^1$-hardness of satisfiability of formulae of TPTL extended with either multiplication by 2 or dense time, we shall encode the value of a counter by a sequence of that length. However, much further work is needed in this proof because the only operation we have on elements of $D$ is equality.

Let $n$ be the number of instructions in $M$. We encode a configuration $\langle l, c_1, c_2 \rangle$ by a sequence of elements of $D$ of the form

$$ddd'd \underbrace{\ldots d' \ldots}_{n} f_1^1 \ldots f_{c_1}^1 eee'e'' f_1^2 \ldots f_{c_2}^2$$

where:

**(i)** the only two pairs of equal consecutive elements are $dd$ and $ee$, and also $f_{c_2}^2$ is distinct from the first element in the encoding of the next configuration,

**(ii)** $e \neq e''$,

**(iii)** after the first 4 elements, there is a sequence of $n$ elements, and only the $l^{\text{th}}$ equals $d'$,

$$\phi_n^{init} \stackrel{\text{def}}{=} \text{start}_d \wedge \mathbf{X}^2 x = \mathbf{X}^4 x \wedge \mathbf{X}^{n+4}(\text{start}_e \wedge \mathbf{X}^4(\text{start}_{d \vee e}))$$

$$\phi_n^{glob} \stackrel{\text{def}}{=} \mathbf{G}(\text{start}_d \Rightarrow \psi_n^1 \wedge \text{start}_e \Rightarrow \psi_n^2)$$

$$\psi_n^1 \stackrel{\text{def}}{=} \overbrace{\left(\bigwedge_{i=1}^{n+3} \mathbf{X}^i x \neq \mathbf{X}^{i+1} x\right)}^{\text{in } dd'd...d'... \text{ any two consecutive values are distinct}} \wedge$$

$$\overbrace{\left(\bigvee_{l=1}^{n} \mathbf{X}^2 x = \mathbf{X}^{l+3} x \wedge \left(\bigwedge_{j=1}^{l-1} \mathbf{X}^2 x \neq \mathbf{X}^{j+3} x \wedge \bigwedge_{j=l+1}^{n} \mathbf{X}^2 x \neq \mathbf{X}^{j+3} x\right)\right)}^{\text{in } ...d'... \text{ exactly one value equals } d'}$$

$$\wedge \overbrace{\mathbf{X}^{n+4}(\psi^{dist} \mathbf{U}\ \text{start}_e)}^{f_1^1...f_{c_1}^1 \text{ mutually distinct}}$$

$$\psi_n^2 \stackrel{\text{def}}{=} \left(\bigwedge_{i=1}^{3} \mathbf{X}^i x \neq \mathbf{X}^{i+1} x\right) \wedge \overbrace{\mathbf{X}^4(\psi^{dist} \mathbf{U}\ \text{start}_d)}^{f_1^2...f_{c_2}^2 \text{ mutually distinct}}$$

$$\psi^{dist} \stackrel{\text{def}}{=} \neg\text{start}_{d \vee e} \wedge \downarrow_{y=x} \mathbf{X}((\neg\text{start}_{d \vee e} \wedge x \neq y) \mathbf{U}\ \text{start}_{d \vee e})$$

Fig. 1.

**(iv)** $f_1^i, \ldots, f_{c_i}^i$ are mutually distinct, for each $i$.

We write $\text{start}_{d \vee e}$ to denote the formula $x = \mathbf{X}^1 x$ stating that the current state is an occurrence of either $dd$ or $ee$. We write $\text{start}_d$ [resp. $\text{start}_e$] to denote the formula $\text{start}_{d \vee e} \wedge x = \mathbf{X}^3 x$ [resp. $\text{start}_{d \vee e} \wedge x \neq \mathbf{X}^3 x$] stating the current state is a first occurrence of $d$ [resp. $e$] in $dd$ [$ee$].

The formula $\phi_M$ is defined as a conjunction

$$\phi_n^{init} \wedge \phi_n^{glob} \wedge \phi_M^1 \wedge \cdots \wedge \phi_M^n \wedge \phi^{rec}$$

where the first two conjuncts state that the model is a concatenation of configuration encodings which satisfy (i)–(iv) above, and that it begins with an encoding of the initial configuration $\langle 1, 0, 0 \rangle$. Their definitions are given in Figure 1.

For any $l \in \{1, \ldots, n\}$, $\phi_M^l$ states that, whenever the model contains an encoding of a configuration $\langle l, c_1, c_2 \rangle$, then the next encoding is of a configuration which is obtained by executing the $l^{\text{th}}$ instruction.

Consider the most complex case: $l : C_2 := C_2 - 1; \text{goto } l' \text{ or goto } l''$. The formula $\phi_M^l$ needs to state that, whenever the location counter is $l$, $C_1$ remains the same, $C_2$ either remains 0 or is decremented, and the next value of the location counter is either $l'$ or $l''$:

$$\chi_{dec}^2 \stackrel{\text{def}}{=} \overbrace{((x = \mathbf{X}^1 x \lor \mathbf{X}^1 x = \mathbf{X}^2 x) \land (\neg \text{start}_e \mathbf{U}(\text{start}_e \land \mathbf{X}^4(x = \mathbf{X}^1 x))))}^{0 \leq C_2 \leq 1 \text{ and the next value of } C_2 \text{ equals } 0} \lor$$

$$\overbrace{(\neg(x = \mathbf{X}^1 x \lor \mathbf{X}^1 x = \mathbf{X}^2 x)}^{C_2 > 1} \land$$

$$\overbrace{(\downarrow_{y=x} \neg \text{start}_e \mathbf{U}(\text{start}_e \land \mathbf{X}^4(\neg \text{start}_d \land x = y))) \land}^{(A)}$$

$$\overbrace{((\neg \mathbf{X}^2 \text{start}_{d \lor e} \land (\downarrow_{y=x} \mathbf{X} \downarrow_{y'=x} (\neg \text{start}_e \mathbf{U}(\text{start}_e \land \mathbf{X}^4(x \neq y \mathbf{U}(x = y \land \mathbf{X}^1 x = y'))))))\mathbf{U} \mathbf{X}^2 \text{start}_{d \lor e}) \land}^{(B)}$$

$$\overbrace{((\mathbf{X}^2 \neg \text{start}_d) \mathbf{U}(\mathbf{X}^2 \text{start}_d \land \downarrow_{y=x} \neg \text{start}_e \mathbf{U}(\text{start}_e \land \mathbf{X}^4(x \neq y \mathbf{U} (x = y \land \neg \text{start}_d \land \mathbf{X}^2 \text{start}_d)))))}^{(C)}$$

Fig. 2.

$$\phi_M^l \stackrel{\text{def}}{=} \mathbf{G}((\text{start}_d \land \mathbf{X}^2 x = \mathbf{X}^{l+3} x) \Rightarrow \\
\mathbf{X}^{n+4}(\chi_{eq}^1 \land (\neg \text{start}_{d \lor e} \mathbf{U}(\text{start}_e \land \\
\mathbf{X}^4(\chi_{dec}^2 \land (\neg \text{start}_{d \lor e} \mathbf{U}(\text{start}_d \land \\
(\mathbf{X}^2 x = \mathbf{X}^{l'+3} x \lor \mathbf{X}^2 x = \mathbf{X}^{l''+3} x)))))))))$$

The formula $\chi_{dec}^2$ given in Figure 2 specifies that, if the current value of $C_2$ is either 0 or 1, then the next value of $C_2$ is 0; and if neither, then the next encoding of the value of $C_2$ equals the current encoding with the last element removed.

The latter is specified as the following conjunction:

**(A)** the first element of the current encoding equals the first element of the next encoding, and

**(B)** for any consecutive pair $y$ and $y'$ of elements in the current encoding such that $y'$ is not the last element, the first occurence of $y$ in the next encoding is followed by $y'$, and

**(C)** the element before the last in the current encoding is the last element in the next encoding.

The formula $\chi_{eq}^1$, which specifies that the value of $C_1$ remains the same, is defined similarly.

Definitions of $\phi_M^l$ for other forms of instruction use the same machinery. For incrementing a counter, it is not necessary to specify that the additional element in the next encoding is distinct from the rest, because that is ensured by $\phi_n^{glob}$.

Finally, $\phi^{rec} \stackrel{\text{def}}{=} \mathbf{GF}(\text{start}_d \wedge \mathbf{X}^2 x = \mathbf{X}^4 x)$ states that the model encodes a recurring computation. $\square$

By Propositions 2 and 3, we have that Theorem 3 can be strengthened by restricting to the fragment of $\text{CLTL}^{\downarrow}(D, =)$ with $|\text{FleVarSet}| = 1$, $|\text{RigVarSet}| = 2$ and such that the flexible variable occurs only in freeze quantifiers of the form $\downarrow_{y=x}$.

By adapting the proof of Theorem 3, the variant of $\text{CLTL}^{\downarrow}(D, =)$ over models which are finite words is also undecidable, more precisely $\Sigma_1^0$-hard through encoding the Halting Problem for 2-counter machines. This should be compared with the undecidability of universality of 1-way nondeterministic register automata [31, Theorem 5.1].

The proof of Theorem 3 can also be modified to yield, for $\text{CLTL}^{\downarrow}(D, =)$ augmented with the past-time operator $\mathbf{U}^{-1}$ ('since') but restricted to 1 rigid variable, $\Sigma_1^1$-hardness over infinite models and $\Sigma_1^0$-hardness over finite models. The sets of values from $D$ which are used to encode counter values do not have to be enumerated in the same order for consecutive configurations, and simpler logical formulae suffice. These results are related to the undecidability of emptiness of 2-way deterministic register automata: see [32, Section 7], [31, Theorem 5.3].

## 5  Related work

In this section, we compare the logic $\text{CLTL}^{\downarrow}(\mathbb{N}, =)$ and the results in this paper with a number of related works in the literature. We show that there is a surprising variety of formalisms which involve the freeze quantifier or related constructs, revealing links among several works which appear unconnected. This confirms that the binding mechanism of the freeze quantifier is fundamental.

**LTL over concrete domains.** Complexity results for Constraint LTL over concrete domains can be found in [16,17,11,18,14] (see also related results for description logics over concrete domains in [33]). Decidability and complexity issues for LTL over Presburger constraints have been studied for instance in [34,22,10,14]. Most decision procedures in the above-mentioned works are automata-based whereas undecidability proofs often rely on an easy encoding of the Halting Problem for 2-counter machines.

LTL over integer periodicity constraints augmented with the freeze quantifier is shown EXPSPACE-complete [14] but CLTL($\mathbb{N}, <, =$) with past-time operator $\mathbf{F}^{-1}$ and $\downarrow$ is undecidable [14].

**Real-time logics.** Similar issues for real-time and modal logics equipped with the freeze quantifier have been considered in [12,35,13,36]. In spite of its rich language of constraints, TPTL model-checking is decidable [12] (discrete version). In this case, decidability is due to the subtle combination of the constraint system and the semantical restrictions (see also versions of metric temporal logics in [37,38]).

The class of logics CLTL$^\downarrow(\mathcal{D})$ defined in this paper is quite general and it is not difficult to show that discrete-time TPTL [12] is exactly the fragment of CLTL$^\downarrow(\mathcal{D})$ where

- $D = \mathbb{N}$ and the only flexible variable is $t$ (time),
- the predicates of $\mathcal{D}$ are

  $$(x \leq c)_{c \in \mathbb{Z}}, (x \leq y + c)_{c \in \mathbb{Z}}, (x \equiv_d c)_{c,d \in \mathbb{N}}, (x \equiv_d y + c)_{c,d \in \mathbb{N}}$$

  where $\equiv_d$ is equality modulo $d$, and
- the formulae are of the form $\mathbf{G}(t \leq \mathbf{X}t) \wedge \mathbf{GF}(t < \mathbf{X}t) \wedge \phi$ with any use of the freeze quantifier being of the form $\downarrow_{x=t}$.

In [12, Theorem 5], $\Sigma_1^1$-hardness of satisfiability for TPTL without the monotonicity condition on time sequences is established. By Propositions 2 and 3, CLTL$^\downarrow(\mathbb{N}, =)$ restricted to one flexible variable can be seen as the fragment of TPTL where there are no atomic propositions, and where the only operation on time is equality. Moreover, it is straightforward to see that Theorem 3 in this paper still holds when satisfiability is restricted to models which contain infinitely many values, which is equivalent to the progress condition when the domain is $\mathbb{N}$. Therefore, a corollary of Theorem 3 is the following strengthening of [12, Theorem 5]: satisfiability for TPTL without the monotonicity condition remains $\Sigma_1^1$-complete even without atomic propositions and with only equality constraints. (The proof of [12, Theorem 5] uses arithmetic on time values.)

**Hybrid, navigation, spatio-temporal, and similar logics.** Hybrid logics (see e.g. [39,40,41]) contain a variable-binding mechanism similar to the freeze quantifier: $\downarrow_x \phi(x)$ holds true iff $\phi(x)$ holds true when the propositional variable $x$ is interpreted as a singleton containing the current state. The downnarrow binder in such hybrid logics records the value of the current state.

Similarly, in temporal logic with forgettable past [26], the effect of the **Now** operator is that the origin of time takes the value of the current state: the states before the current state are forgotten. Identical mechanisms are used in navigation logics for object structures, see e.g. [42] and in half-order dynamic temporal logics interpreted over traces from sequence diagrams [43].

In the context of spatio-temporal logics, Wolter and Zakharyaschev [16, Section 7] advocate the need to consider operators expressing constraints of the form $\bigwedge_{i \in \mathbb{N}} R(x, \mathbf{X}^i y)$ and $\bigvee_{i \in \mathbb{N}} R(x, \mathbf{X}^i y)$. They are simple to express in CLTL$^\downarrow(\mathcal{D})$, as $\downarrow_{x'=x} \mathbf{G} R(x', y)$ and $\downarrow_{x'=x} \mathbf{F} R(x', y)$. These formulae are in the flat fragment: see Section 3.2.

**Quantified propositional temporal logic with repeating.** The models of Quantified Propositional Temporal Logic with Repeating (also known as RQPTL) introduced in [44] can be encoded by CLTL$^\downarrow(\mathbb{N}, =)$ formulae, unlike the second-order quantification in the language. Such models are pairs of maps $\langle \mu : \mathbb{N} \to S, \pi : S \to 2^{\mathrm{AP}} \rangle$ where $S$ is an arbitrary set (of states). A possible encoding is by treating $\mu$ as the interpretation of a distinguished flexible variable, and using the freeze quantifier to specify that, whenever $\mu(i) = \mu(j)$, any propositional variable has the same values at time points $i$ and $j$. (See Section 2.4 regarding encodings of propositional variables.)

On the other hand, the variant logic RHLTL$^n$ [44, Section 4] can be shown equivalent to CLTL$^\downarrow(\mathbb{N}, =)$ with one flexible variable and $n$ rigid variables, except that RHLTL$^n$ does not have the $\mathbf{U}$ operator but has $\mathbf{F}$ and the past-time operators $\mathbf{F}^{-1}$ and $\mathbf{X}^{-1}$. Theorem 3 in this paper and $\Sigma_1^1$-hardness of RHLTL$^2$ [44, Corollary 1] are therefore complementary results.

**Predicate λ-abstraction.** A number of decidability and undecidability results for half-order modal logics (to be compared with [35]) are presented in [45]. The half-order aspect of such logics is due to a predicate λ-abstraction mechanism, which solves the famous problem of interpreting constants in modal logic. Even though this construct is essentially the same as the freeze quantifier, apparently there have been no cross-references between the literature dealing with predicate λ-abstraction (e.g. [45,15]) and that dealing with the freeze quantifier (e.g. [35,12,14,1]). However, several undecidability results for LTL-like logics with predicate λ-abstraction have recently been obtained in [15], independently and concurrently with [1]. The most related to Theorem 3 in this paper are $\Sigma_1^1$-hardness results for the following logics:

**(I)** LTL$_{\lambda=}$ with temporal operators $\mathbf{X}$ and $\mathbf{U}$, and with 3 rigid variables;
**(II)** LTL$_\lambda$ with temporal operators $\mathbf{X}$ and $\mathbf{U}$, and with countably infinitely many unary predicate symbols (but no equality).

Remarkably, LTL$_{\lambda=}$ is essentially the same as CLTL$^{\downarrow}$($\mathbb{N}$, =). The proofs of (I) in [15] and of Theorem 3 above reduce from the same $\Sigma_1^1$-hard problem. However, the encodings are different, enabling Theorem 3 to be sharper by restricting to 1 flexible and 2 rigid variables.

An interesting discussion of applications to dynamic systems with resources, like communication protocols for mobile agents, can also be found in [15].

**Monodic first-order temporal logics.** Since freeze quantification is first-order quantification over a singleton set, the freeze quantifier can be expressed in first-order temporal logics [46,47,48,49]. Indeed, CLTL$^{\downarrow}$($\mathbb{N}$, =) satisfiability can be reduced to first-order temporal logic $\mathcal{TL}$ satisfiability over the linear structure $\langle \mathbb{N}, < \rangle$ (the latter logic was introduced in [49, Chapter 11]). To each flexible variable $x$ one associates a monadic predicate symbol $P_x$ in such a way that $P_x$ is interpreted as the singleton set containing the value of $x$. A formula of the form $\downarrow_{x'=\mathbf{X}x} \phi$ is then translated to $\exists x' \ \mathbf{X}P_x(x') \wedge \phi'$ where $\phi'$ is the translation of $\phi$. The translation is homomorphic for Boolean and temporal operators, whereas for instance $y = \mathbf{X}z$ with $y, z \in$ FleVarSet is translated into $\exists x \ P_y(x) \wedge \mathbf{X}P_z(x)$. One needs also to be able to express that at every state $P_x$ is interpreted by a singleton, which can be encoded by the formula $\mathbf{G}(\exists z \ P_x(z) \ \wedge \ \forall z, z'(P_x(z) \wedge P_x(z') \Rightarrow z = z'))$.

Consider the fragment of CLTL$^{\downarrow}$($\mathbb{N}$, =) with |RigVarSet| = 1. It is easy to check that its translation is contained in the *monodic* fragment of $\mathcal{TL}$ with equality, and with only two individual variables and monadic predicate symbols. We recall that in the monodic fragment, any temporal subformula (i.e. whose outermost construct is a temporal operator) must have at most one free individual variable. Even though monodic $\mathcal{TL}$ over $\langle \mathbb{N}, < \rangle$ is decidable [50], its extension with equality is not [47], even with the above restrictions [46].

**Logics and automata for data languages.** In [51,52], data languages are defined as sets of finite data words in $(\Sigma \times D)^*$ where $\Sigma$ is a finite alphabet and $D$ is an infinite domain (generalising the concept of timed languages), and automata which recognise data languages are introduced. The latter are related to register and pebble automata for strings over infinite alphabets (e.g. [31]).

First-order logic over finite data word models is considered in [53], with motivations stemming from query languages for semistructured data. More precisely, the carrier of a model is the set of positions in a data word, there are no function symbols, the unary predicates correspond to elements of $\Sigma$, and there are binary predicates $<, +1$, as well as $\sim$ which is interpreted as equality of elements of $D$ at given positions. FO$^k(\sim, <, +1)$ denotes such a logic with $k$ variables. The main result of [53] is that satisfiability of $\text{FO}^2(\sim,<,+1)$ is decidable, by a doubly exponential-time reduction to nonemptiness of multi-counter automata. (The latter problem is decidable, but there is no known elementary upper bound.)

The following variant of $\text{CLTL}^{\downarrow}(D,=)$ has models which are words over $\Sigma \times D$: there is one flexible variable $x$ which takes values in $D$, plus one flexible variable $l$ which takes values in $\Sigma$ and on which freeze quantification cannot be used, but to which unary predicates $P_a$ for equality testing with $a \in \Sigma$ can be applied. Interestingly, that logic with infinite $D$ and 1 rigid variable is incomparable with $\text{FO}^2(\sim,<,+1)$. In one direction, $\text{FO}^2(\sim,<,+1)$ cannot express the $\mathbf{U}$ operator, and also not formulae of the form $\downarrow_{y=x} \phi$ where $y$ occurs in $\phi$ under two or more temporal operators. In the other direction, $\text{FO}^2(\sim,<,+1)$ can express past-time operators such as $\mathbf{F^{-1}}$.

## 6 Conclusion

We have shown that adding the freeze quantifier to $\text{CLTL}(\mathcal{D})$ leads to undecidability as soon as the underlying domain is infinite and the equality predicate is part of $\mathcal{D}$. As illustrated in the paper, in most related work dealing with undecidable logics having a binding mechanism similar to freeze quantification, either past-time operators can be encoded or constraints richer than equality are available.

The logic $\text{CLTL}^{\downarrow}(\mathcal{D})$ is EXPSPACE-complete for most of finite domains $\mathcal{D}$. In order to design a specification language over infinite domains with LTL temporal operators and the freeze quantifier that admits a decidable model-checking problem, syntactic restrictions could be a reasonable solution. The existence of a logarithmic-space reduction from the flat fragment of $\text{CLTL}^{\downarrow}(\mathcal{D})$ into $\text{CLTL}(\mathcal{D})$ when the equality predicate is present leads us to believe that the flatness criterion is most relevant here.

As we have seen, the following fragments/variants of $\text{CLTL}^{\downarrow}(D,=)$ with infinite $D$ and $|\mathsf{FleVarSet}| = 1$ are $\Sigma_1^1$-hard:

- the temporal operators are $\mathbf{X}$ and $\mathbf{U}$, and $|\mathsf{RigVarSet}| = 2$;
- the temporal operators are $\mathbf{X}$, $\mathbf{U}$ and $\mathbf{U^{-1}}$, and $|\mathsf{RigVarSet}| = 1$;
- the temporal operators are $\mathbf{X}$, $\mathbf{X^{-1}}$, $\mathbf{F}$ and $\mathbf{F^{-1}}$, and $|\mathsf{RigVarSet}| = 2$;

It is open whether the intersections of these fragments are decidable.

Other open problems include:

- decidability in the presence of semantic restrictions such as reversal bound-

edness [5] of a flexible variable;
- decidability over infinite domains without equality (and where equality is not definable by other predicates), such as $\langle \{0,1\}^*, < \rangle$ with $<$ being either the strict prefix relation or the strict subword relation.

**Acknowledgements.** We are grateful to Deepak D'Souza, Claire David, Anca Muscholl and Luc Segoufin for helpful discussions, and to Frank Wolter for having directed us to related work.